\documentstyle[aaspp4,11pt,psfig]{article}   

\newcommand {\be} {\begin{equation}}
\newcommand {\ee} {\end{equation}}

\def\nss{neutron stars}

\begin{document}

\title{Constraints on Thermal Emission Models of Anomalous X-ray Pulsars}
 
 \author{Feryal \"Ozel\altaffilmark{1}, Dimitrios
 Psaltis\altaffilmark{2}, and Victoria M.\ Kaspi\altaffilmark{3,2}}
 \altaffiltext{1}{Harvard-Smithsonian Center for Astrophysics and
 Physics Department, Harvard University, 60 Garden St., Cambridge, MA
 02138; fozel@cfa.harvard.edu } \altaffiltext{2} {Center for Space
 Research, Massachusetts Institute of Technology, Cambridge, MA 02139;
 demetris@space.mit.edu } \altaffiltext{3}{Department of Physics,
 Rutherford Physics Building, McGill University, 3000 University
 Street, Montreal, Quebec, H3A 2T8, Canada; vkaspi@physics.mcgill.ca}

 \begin{abstract} 
 
 Thermal emission from the surface of an ultramagnetic neutron star is
 believed to contribute significantly to the soft X-ray flux of the
 Anomalous X-ray Pulsars (AXPs). We compare the detailed predictions
 of models of the surface emission from a magnetar to the observed
 spectral and variability properties of AXPs. In particular, we focus
 on the combination of their luminosities and energy-dependent pulsed
 fractions. We use the results of recent calculations for strongly
 magnetized atmospheres in radiative equilibrium to obtain the angle-
 and energy-dependence of the surface emission. We also include in our
 calculations the significant effects of general relativistic photon
 transport to an observer at infinity as well as the effects of
 interstellar extinction. We find that the combination of the large
 pulsed fractions and the high inferred luminosities of AXPs cannot be
 accounted for by surface emission from a magnetar with two antipodal
 hot regions or a temperature distribution characteristic of a
 magnetic dipole. This result is robust for reasonable neutron star
 radii, for the range of magnetic field strengths inferred from the
 observed spin down rates, and for surface temperatures consistent
 with the spectral properties of AXPs. Models with a single hot
 emitting region can reproduce the observations, provided that the
 distance to one of the sources is $\sim 30\%$ less than the current
 best estimate, and allowing for systematic uncertainties in the
 spectral fit of a second source. Finally, the thermal emission models
 with antipodal emission geometry predict a characteristic strong
 increase of the pulsed fraction with photon energy, which is apparently
 inconsistent with the current data. The energy-dependence of the
 pulsed fraction in the models with one hot region shows a wider range
 of behavior and can be consistent with the existing data. Upcoming
 high-resolution observations with {\em Chandra} and {\em XMM-Newton}
 will provide a conclusive test.
\end{abstract}
 
\keywords{radiation mechanisms:thermal --- stars:magnetic fields --- 
 stars:neutron --- X-rays:stars}

\newpage  
\section{INTRODUCTION}

Anomalous X-ray Pulsars (AXPs), first identified as a new class by
Mereghetti \& Stella (1995; see also Hellier 1994), are bright
($L_{\rm X} = 10^{34}-10^{36} {\rm erg~s^{-1}}$) X-ray sources with
soft spectra. They have periods clustered between 6$-$12~s and no
detectable radio counterparts. Of the five sources and one candidate
known to date, three are found in supernova remnants (Gaensler et al.\
2001). These associations as well as the low galactic latitudes of the
AXPs imply that they are a young population of neutron stars. Their
large period derivatives ($\dot P \sim 10^{-11} \rm{s}\,\rm{s}^{-1}$),
together with the constraints on the presence of companion stars set
by Doppler delay measurements and the tight limits on the optical
emission from an accretion disk (see, e.g., Mereghetti, Israel, \&
Stella 1998; Hulleman, van Kerkwijk, \& Kulkarni 2000) led to the
suggestion that AXPs are isolated \nss\ with very large magnetic field
strengths ($B\sim 10^{14}-10^{15}$~G; Thompson \& Duncan 1996).

In a group of models based on this hypothesis, collectively referred
to as magnetar models, a large fraction of the X-ray emission of AXPs
is powered by the decay of this strong magnetic field (see, e.g.,
Thompson \& Duncan 1996), by neutron star cooling (Heyl \& Hernquist
1998), by crustal fracturing under large magnetic torques (Duncan
1998), or even by isolated regions of intense magnetic activity on the
neutron star surface. In all these cases, the heat released in the
crust emerges as thermal radiation through the neutron star
atmosphere, which has an anisotropic temperature distribution due to
the presence of the strong magnetic field. In addition, magnetospheric
processes may have a significant contribution to the total emission
and have been suggested as the origin of hard tails observed in the
spectra of AXPs. In the present paper, we focus on the surface
emission, and compare the detailed predictions of these models to the
properties of thermal-like emission observed in AXPs (see \S2.2).

The possibility of constraining models of AXPs based on the observed
properties of their pulse profiles was recently pointed out by DeDeo,
Psaltis, \& Narayan (2001). A very prominent characteristic of AXPs is
that their X-ray fluxes can be highly modulated during a pulse cycle.
Correspondingly, they have a wide range of pulsed fractions, extending
from $15\%$ to as high as $80\%$ in the $1-10$~keV photon-energy range
(see, e.g., Pivovaroff, Kaspi, \& Camilo 2000). Taking into account
the suppression of pulsed fractions caused by general relativistic
light bending around the neutron star surface, DeDeo et al.\ (2001)
show that the high end of this range of pulsed fractions can be
produced only if the surface emission is localized and highly
beamed. This calculation, however, is carried out for bolometric
pulsed fractions and relies on simple, radially peaked mathematical
functions for describing the beaming of the emerging radiation.  Using
a similar description, Perna, Heyl, \& Hernquist (2000) argue that the
effects of interstellar extinction might significantly increase the
bolometric pulsed fractions. We discuss these effects in \S 3.4.

Simultaneous consideration of the variability properties and the X-ray
luminosities of AXPs provides an even more powerful diagnostic of the
underlying emission mechanism (Psaltis, \"Ozel, \& DeDeo 2000). This
is because, for a thermally emitting star of a given effective
temperature, a small emitting area minimizes the pulse-phase averaged
luminosity, while maximizing the pulsed fraction. This was
demonstrated in the case of blackbody radiation that emerges
isotropically from the neutron-star surface. In the present work, we
use this effect as a test of the thermal emission models proposed for
the AXPs, because they are simultaneously bright and can reach high
pulsed fractions.  Furthermore, we propose to use as another test of
emission models the energy-dependence of the pulsed fraction, which
carries characteristic signatures of the underlying emission
mechanism. Applying the above diagnostics to AXPs as well as going
beyond the study of purely bolometric pulsed fractions requires the
calculation of the photon-energy and angle dependences of the
radiation emerging from an ultramagnetized neutron star atmosphere.

Models of neutron-star atmospheres in radiative equilibrium have
recently been calculated for the $\sim 10^{14}-10^{15}$~G magnetic
field strengths required by the magnetar models of AXPs (\"Ozel 2001a;
Ho \& Lai 2001). These calculations demonstrate that the emerging
spectra are broader in shape than pure blackbodies and have color
temperatures that are higher than the effective temperatures of the
atmospheres (see also Zane et al.\ 2001 and \S3.2 for a
discussion). \"Ozel (2001a) also shows that taking into account vacuum
polarization effects leads to the generation of a power-law tail at
high photon energies for high magnetic field strengths.  Moreover,
angle-dependent radiative transfer calculations show that the beaming
of the emerging radiation depends strongly on photon energy, is
largely non-radial, and, therefore, cannot be described by any of the
simple beaming functions used in previous studies (\"Ozel 2001a). As a
result, general relativistic light bending may actually lead to an
increase of the pulsed fraction for such non-radially peaked radiation
patterns (\"Ozel 2001b). This thus affects the generality of the
results of DeDeo et al.\ (2001).

In this paper, we use the results of detailed calculations of magnetar
atmospheres (\"Ozel 2001a, 2001b) in order to compare the predicted
spectral and variability properties of thermally emitting magnetar
models to a synthesis of all available data on AXPs. In particular, we
apply two diagnostic tests relevant for thermal emission models (\S 4)
to {\em ASCA\/} and {\em BeppoSAX\/} observations of AXPs in order to
assess the viability of such models.  We begin by reviewing the
observed properties of AXPs, focusing on their pulsed fractions,
thermal luminosities, and distance estimates.

\section{THE DATA}

In this section, we discuss the observed properties of AXPs,
summarized in Figure~\ref{Fig:bbody}, and Tables~$1-4$ (see also
Mereghetti 2001 for a review).  The list of 5 known AXPs, along with
their periods, period derivatives, and inferred dipole magnetic field
strengths are given in Table~1. We consider in detail their pulse
morphologies, pulsed fractions, spectra, and the distance estimates to
these sources, all of which play an important role in the proposed
diagnostic tests discussed in detail in \S3 and \S4.  Because three
AXPs are inside supernova remnants (Gaensler et al.\ 2001), accurate
measurements of their pulsed fractions and X-ray spectra require the
use of imaging telescopes to remove the contamination from the
remnants. For this reason we focus our discussion on observations
carried out with imaging telescopes.

\subsection{Pulse Profiles and Pulsed Fractions}

\begin{figure}[t]
 \centerline{ \psfig{file=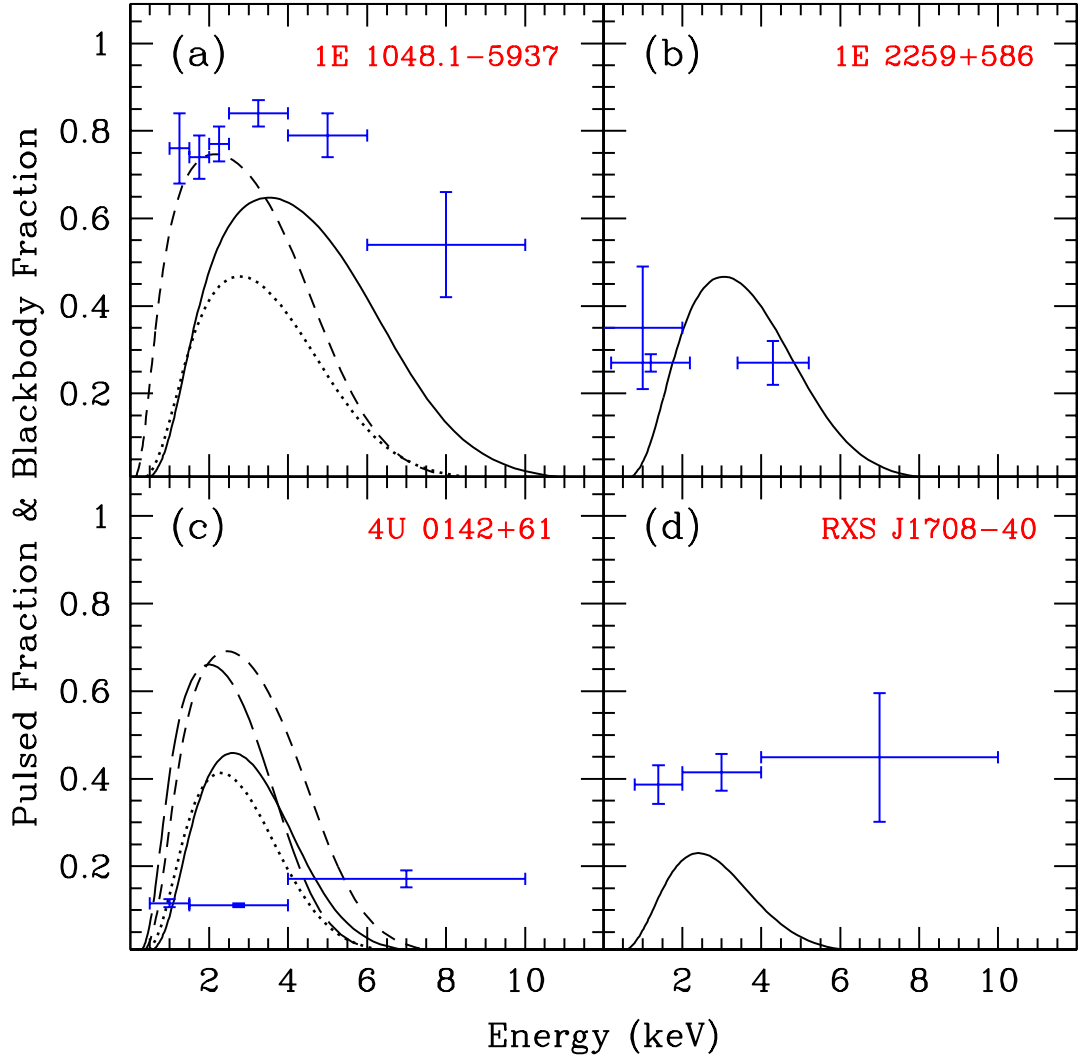,angle=0,width=15truecm} }
\figcaption[]{The observed pulsed fraction (error bars) and the fractional
contribution of the blackbody component to the X-ray spectrum (lines)
as a function of photon energy, for four AXPs. The multiple lines in
each panel correspond to fits to different data sets; the references
and details are given in Table~3. \label{Fig:bbody}}
\end{figure}

The pulse profiles of AXPs are characterized by large duty cycles (see
Gavriil \& Kaspi 2001 and references therein) and pulsed fractions in
the range $15\%$ to $80\%$ that depend weakly on photon energy (see
Table~\ref{Tab:PF} and Fig.~\ref{Fig:bbody}).  Monthly monitoring of
five AXPs over three years with the {\em Rossi X-ray Timing
Explorer\/} showed no detectable variability of their pulse shapes
(Gavriil \& Kaspi 2001), indicating that the pulse profiles of AXPs
are largely stable (but see Iwasawa, Koyama, \& Halpern 1992 for
possible short-duration changes in the pulse profile of
1E~2259$+$586).

AXPs show a varying degree of structure in their pulse profiles. With
the exception of 1E~1048.1$-$5937 which shows a highly sinusoidal
pulse (e.g., Oosterbroek et al.\ 1998), the profiles of the other AXPs
show multiple peaks in a pulse cycle and significant power in the
second harmonic (Gavriil \& Kaspi 2001). We can make use of the
structure of the pulse profiles to infer the emission geometry. In the
framework of the thermally emitting magnetar model, with its
non-radial beaming pattern, this structure can be accounted for either
by the presence of two antipodal emission regions or by a single
emission region, depending on the observer angle (\"Ozel 2001b).  We
consider both of these cases in \S4.

To characterize pulse variability of AXPs, we consider the
peak-to-peak changes of the X-ray flux during a pulse cycle, as
measured by the pulsed fraction defined as
\be
PF\equiv \frac{F_{\rm max}-F_{\rm min}}{F_{\rm max}+F_{\rm min}}\;.
\ee
Here, $F_{\rm max}$ and $F_{\rm min}$ are the maximum and minimum
values of the observed photon flux. We now consider the pulsed
fraction and its energy dependence for each source separately.

For 1E~1048.1$-$5937, the weak energy dependence of the pulsed
fraction was determined from {\em BeppoSAX\/} observations by
Oosterbroek et al.\ (1998). The pulsed fraction was found to be $\sim
75-85\%$ in six energy bands (see Fig.~\ref{Fig:bbody}a). Similar
results were also obtained by Corbet \& Mihara (1997) and Paul et al.\
(2000) with {\em ASCA\/} observations.

For 1E~2259$+$586, which lies in the supernova remnant CTB~109, pulse
profiles at different photon energies have been published mostly for
observations made with non-imaging detectors ({\em Tenma:\/} Koyama, Hoshi,
\& Nagase 1987; {\em EXOSAT:\/} Hanson et al.\ 1988, Morini et al.\ 1988;
{\em Ginga:\/} Koyama et al.\ 1989). The contamination from the
remnant emission at low photon energies leads to an apparent increase
of the pulsed fraction with energy (as, e.g., reported by Koyama et
al.\ 1989). Hanson et al.\ (1988) compared the $0.35\pm 0.14$ pulsed
fraction measured using the imaging telescope (LE) onboard {\em
EXOSAT\/} at low energies ($\lesssim 2$~keV), with the $0.27\pm 0.05$
value measured using the non-imaging detector (ME) at higher photon
energies ($2.5-3.5$~keV, which are less contaminated by the remnant)
and suggested that in this source the pulsed fraction depends very
weakly on photon energy. This is also consistent with the results of
Rho \& Petre (1997) who published the pulse profile observed with {\em
ROSAT/PSPC} in the $0.2-2.2$~keV range (the exact energy bounds are
not quoted in the paper), from which we infer a pulsed fraction of
$0.27\pm 0.02$ (Fig.~\ref{Fig:bbody}b).

Analysis of the pulse profiles of 4U~0142$+$61 at different photon
energies yields consistent pulsed fractions between {\em BeppoSAX\/}
(Israel et al.\ 1999) and {\em ASCA\/} observations (White et al.\
1996; Paul et al.\ 2000). We use the pulse profiles of Israel et al.\
(1999) to infer the pulsed fractions in three energy bands, which
yield $12\%$ in the two lower and $17\%$ in the highest bands
(Fig.~\ref{Fig:bbody}c).

For 1RXS~J170849$-$4009 (hereafter referred to as RXS~J1708$-$40 for
simplicity), we infer the pulsed fraction in three energy bands using
the pulse profiles from {\em ASCA\/} observations reported by Sugizaki
et al.\ (1997). The results are shown in Figure~\ref{Fig:bbody}d.

Finally, the low observed countrates from 1E~1841$-$045 do not allow
for a comparison of pulsed fractions at different photon
energies. Vasisht \& Gotthelf (1997) report that the pulsed fraction
in the ASCA/GIS energy band ($\sim 1-10$~keV) is roughly $15\%$ after
subtracting the contribution of the supernova remnant to the flux.
(Note that the actual value for the pulsed fraction quoted in Vasisht 
\& Gotthelf 1997 is $30\%$ because of a factor of 2 difference in the
definition of this quantity adopted by these authors.)

\subsection{Spectral Fits}

The observed X-ray spectra of AXPs are soft and suggest the presence
of thermal emission originating from the surfaces of the neutron
stars. However, in most cases, a single blackbody component is
insufficient to characterize the spectrum and the residual hard excess
is typically modeled with the addition of a steep power law (see
Table~\ref{Tab:BB} and Mereghetti 2001 for a review). Although a
fraction of this hard excess may be due to the effects of the neutron
star atmosphere, the power-law tail has often been assumed to have a
separate, possibly magnetospheric, origin. The blackbody contribution
to the observed flux is large at moderate ($\sim 1-5$~keV) photon
energies, ranging between $\sim 40-80\%$ (see Fig.~\ref{Fig:bbody}).

Recently, Perna et al.\ (2001) fit the spectra of AXPs with a model of
surface emission from a strongly magnetic neutron star with a dipole
field geometry (Heyl \& Hernquist 1998).  They argue that the fits
still require the addition of a power-law component, although we note
that its statistical significance is weak (at the $\sim 1 \sigma$
level) given the large uncertainty in the reported power-law
normalizations (see Table~2 of Perna et al.\ 2001). It should also be
noted that the atmosphere model (Heyl \& Hernquist 1998) employed by
the authors makes use of Rosseland mean opacities which is problematic
because such an averaging, dominated by one of the photon polarization
modes of an ultramagnetized plasma, fails to capture the important
polarization and angle dependence of the magnetic opacities.
Therefore, the resulting radial temperature profiles do not show the
characteristic plateaus at multiple depths in the atmosphere and can
be an order of magnitude too high at the neutron star surface, giving
rise to incorrect spectra (Shibanov et al.\ 1992; \"Ozel 2001a). The
requirement for a power-law component in connection to more detailed
models of surface emission has not yet been explored (also see below).

Regarding such spectral decompositions, we first note that the weak
dependence of the pulsed fraction on photon energy, together with the
fact that the blackbody contribution is highly variable with photon
energy (Fig.~\ref{Fig:bbody}), with a prominent peak at $\sim 3$~keV,
is problematic for any model that consists of two distinct thermal and
non-thermal emission mechanisms of different origins (see also
Oosterbroek et al.\ 1998). For example, for the BeppoSAX observations
of 1E~1048.1$-$5937, the blackbody contribution is $\approx 10\%$ at
$\sim 1$~keV and $60\%$ at $\sim 3$~keV, while the pulsed fraction
remains constant at $\sim 80\%$. Therefore, to explain the weak
energy-dependence of the pulsed fraction would require the two
components to be strongly coupled in a contrived way, or to be pulsed
with the same nearly constant pulsed fraction and without a
significant phase lag.

In either case, since the thermal surface emission, as inferred from
the large blackbody contribution, dominates the total flux at least at
$\sim 1-5$~keV for most AXPs, the surface emission itself needs to be
strongly pulsed in this energy range, given the observed pulsed
fractions. This is true even if the inferred power-law component
dominates the luminosity measured over a much wider energy
range. Therefore, the recent suggestion that the high pulsed fractions
may be due to the power-law components alone, based on the large
contribution of the power law to the flux over a wide energy range
(Perna et al.\ 2001), is not supported by the energy dependences of
the pulsed fractions and of the two spectral components. (Note also
that the inferred power-law contribution to the $0.7-10$~keV flux for
all AXPs found by Perna et al.\ 2001 suffers from large formal
uncertainties).

Recent theoretical models of strongly magnetic neutron-star
atmospheres have shown that the emerging spectra are significantly
broader than a simple blackbody and, depending on the magnetic field
strength, can exhibit power-law tails (\"Ozel 2001a). This suggests
that the surface emission alone may be able to account for the entire
observed spectra, including the power-law tails. Correspondingly,
simple blackbody fits may have caused an overestimate of the inferred
power-law components. Therefore, within the framework of the thermally
emitting magnetar models, the blackbody component of the spectral
decomposition provides only a lower limit to the thermal contribution
from the neutron star surface.

In this paper, we use the decomposition of the observed AXP spectra
into two components only as a mathematical expression that describes
their shapes. We focus on the fluxes of the blackbody components as
lower limits to the emission from the stellar surfaces.  We also use
their measured temperatures ($T_{\rm BB}$ in Table~3) for a direct
comparison of the magnetar models to the existing fits, by setting
$T_{\rm c}^\infty = T_{\rm BB}$, where $T_{\rm c}^\infty$ is the color
temperature of the thermal emission spectrum measured at infinity.

\subsection{Distance Estimates}

AXPs have been discovered in a wide range of galactic longitudes and
with a distribution clustered close to the galactic plane.  The
distances to these sources are uncertain, primarily due to the lack of
counterparts for most of them. Various techniques have been employed,
involving the properties of the compact objects in the X-rays and, in
some cases, those of the associated supernova remnants. Because one of
our diagnostic tools (see \S4 and Fig.~\ref{Fig:2spot}) depends
sensitively on the inferred source luminosities, hence distances,
we review here in detail the current best distance estimates and
assess their uncertainties (see also Table~\ref{Tab:L}).

The AXP 1E~1048.1$-$5937 lies in the direction of the Carina nebula
(Seward, Charles, \& Smale 1986). The spectral fits for this source
require an equivalent neutral hydrogen column density in the range
$N_{\rm H}=(0.5-1.6)\times 10^{22}$~cm$^{-2}$ (see
Table~\ref{Tab:BB}).  The values of $N_{\rm H}\gtrsim 1.5\times
10^{22}$~cm$^{-2}$ that resulted from fitting a simple power-law
spectral model to low-resolution data (Seward et al.\ 1986) are not
favored by subsequent {\em ASCA\/} and {\em BeppoSAX\/} observations.
The lower end of this range implies an optical extinction of $A_{\rm
V}\gtrsim 2$~mag (Gorenstein 1975; Predehl \& Schmitt 1995),
comparable to or exceeding those of stars ($A_{\rm V}\sim 1-3$) that
lie within the Carina nebula (e.g., Turner \& Moffat 1980). This
implies that the source is within or behind the nebula and, therefore,
has a minimum distance of 2.7~kpc, as suggested by Seward et al.\
(1986). We have also verified this conclusion by comparing the
interstellar extinction measurements for other X-ray sources in the
Carina nebula (Corcoran, M. private communication) to the value
inferred for 1E~1048.1$-$5937, reaching consistent results. We
therefore adopt the value of 2.7~kpc as a conservative lower limit to
the distance. (Note that the 10~kpc distance used by Perna et al.\
2001 has been calculated by Mereghetti \& Stella 1995 using
spin-equilibrium arguments for an accreting $10^{11}$~G neutron star
and is, therefore, inconsistent to use in connection with a magnetar
model.)

The distance estimate to 1E~2259$+$586 is linked to the distance
measurements to the supernova remnant CTB~109 (G109.1$-$1.0), which is
likely to be associated with the pulsar (Fahlman \& Gregory 1981; see
also Gaensler et al.\ 2001). This association is based on the spatial
coincidence of the two sources and is supported by the similar X-ray
column density towards the pulsar and remnant. Several independent
measurements exist for the distance towards the supernova remnant. The
strongest (lower) limit of $\simeq 4-5$~kpc is based on
spectrophotometric distances to stars in HII regions within a
molecular cloud that appears to be either in contact with or in front
of CTB~109 (Crampton, Georgelin, \& Georgelin 1978; Tatematsu et al.\
1987). Kinematic 21-cm distance measurements place the remnant as well
as the molecular cloud at a distance of $\simeq 7$~kpc. The
discrepancy with the spectrophotometric distance is likely to be
caused by the assumption of circular orbits in the kinematic distance
analysis (Braun \& Strom 1986). Finally, the uncertain $\Sigma-D$
relation for supernova remnants yields a distance of $\simeq 5.6$~kpc
for CTB~109, though with significant uncertainty (Hughes et al.\
1984). These distance measurements are discussed in more detail in
Wang et al.\ (1992) and Hulleman et al.\ (2000b). Here, we adopt the
conservative range of $4-7$~kpc for the distance to 1E~2259$+$586.

The distance estimates to 4U~0142$+$61, which lies at $l=129^\circ$,
rely on the values of the column density inferred from spectral
fits. Different studies with {\em ASCA\/} and {\em BeppoSAX\/} yield
consistent values; they find $N_{\rm H}=(0.95-1.17)\times
10^{22}$~cm$^{-2}$, which implies $4-5.5$~mag of optical
extinction. Comparing this value to the extinction measured for stars
in two open clusters along the same direction ($2.4-2.7$~mag), Israel,
Mereghetti, \& Stella (1994) concluded that the distance to the AXP is
larger than $2.4-2.7$~kpc, which is determined independently for the
clusters. However, Israel et al.\ (1994) also noted that the distance
to the clusters may not imply a lower limit for the distance to the
AXP, because the line-of-sight to 4U~0142$+$61 crosses the edge of a
local ($\lesssim 1$~kpc) molecular cloud.  Including or neglecting the
contribution of the molecular cloud to the extinction results in two
different lower limits for the distance to the source, both of which
we consider in our calculations.

The source RXS~J1708$-$40 is another AXP for which the distance
estimate relies on the column density inferred from X-ray extinction.
The range of column densities $(1.4-1.8)\times 10^{22}$~cm$^{-2}$
obtained from spectral fits (Sugizaki et al.\ 1997) agrees well with
the total atomic and molecular H column densities obtained using 21-cm
and CO observations in the direction of the source, which lies towards
the Galactic Center. Given that the latter column densities correspond
to the contribution from the entire galaxy along the line of sight,
Sugizaki et al.\ (1997) concluded that the source cannot lie at a
distance much closer than the Galactic Center. We explored the
possibility of local extinction along the line-of-sight to the pulsar
contributing to its column density and implying a smaller distance. We
used W3NH (Dickey \& Lockman 1990,
http://heasarc.gsfc.nasa.gov/cgi-bin/Tools/w3nh/w3nh.pl) and found
that the above values are typical ($[1.2-1.8]\times
10^{22}$~cm$^{-2}$) for lines of sight within a few degrees of the
direction to the pulsar. Therefore, we choose a fiducial distance of
8~kpc for this source.

Finally, a robust determination of the distance to 1E~1841$-$045 comes
from studies of 21-cm absorption towards the supernova remnant Kes~73
(G27.4$+$0.0) associated with the pulsar (Sanbonmatsu \& Helfand
1992).  This yields a kinematic distance of $5.7-8.5$~kpc, where the
effects of non-circular orbits have been taken into account in
determining this range.

\section{PROPERTIES OF THERMAL EMISSION FROM A MAGNETAR}

In magnetar models, thermal emission is expected to arise when heat,
originating from the neutron-star core or released in the crust, is
transported to the surface through the stellar atmosphere. These
surface layers, therefore, shape the observable characteristics of the
neutron star. In this section we explore the spectral and variability
properties of the surface emission predicted by magnetar models
(\"Ozel 2001a,b). In particular, we focus on phase-averaged spectra as
well as on bolometric and photon-energy-dependent pulsed fractions.

\subsection{Description of the Model}

We use the radiative equilibrium atmosphere models described in \"Ozel
(2001a) to calculate the beaming and energy dependence of the
radiation emerging from the surface of a strongly magnetic neutron
star. We carry out the radiative transfer calculations for a fully
ionized, plane-parallel H atmosphere. We take into account
conservative scattering and absorption by free electrons as well as
the effects of virtual pairs present in the magnetic vacuum.  We also
assume that the magnetic field $B$ is orthogonal to the surface. Note
that our diagnostics employ the maximal pulsed fraction from a
thermally emitting neutron star, which is obtained when the magnetic
field is taken to be normal to the surface everywhere on the neutron
star (this is for calculating the beaming of radiation only; we allow
for different magnetic field geometries when calculating the emerging
flux). This is because, in a magnetic dipole geometry, the angle of
the magnetic vector to the surface normal increases from the magnetic
pole to the equator, thus spreading the observable radiation over a
larger fraction of the pulse cycle and suppressing the variability.

We assume a temperature anisotropy on the surface of the neutron star
with hot emitting regions and cold regions that produce negligible
radiation. We specify the flux emitted from the hot regions by an
effective temperature $T_{\rm eff}$, which determines the flux through
a radiative equilibrium atmosphere through $F = \sigma T^4_{\rm eff}$,
where $\sigma$ is the Stefan-Boltzmann constant. We consider one as
well as two identical antipodal hot caps, motivated by the pulse
profiles of AXPs (see \S 2 and \"Ozel 2001b). The temperature in this
case is assumed not to vary across the hot emitting regions, the
angular sizes of which are denoted by $\rho$. In addition, we consider
a second temperature distribution on the surface that is suggested for
a cooling neutron star with a strong dipole magnetic field (Heyl \&
Hernquist 1998). In this case, the variation of the effective
temperature from the magnetic pole to the equator is given by
\be 
T_{\rm eff}(\theta) = T_p
\left[\frac{\cos^2\theta}{(3\cos^2\theta+1)^{0.8}}\right]^{1/4}, 
\label{eq:hh}
\ee
where $T_p$ is the temperature of the magnetic pole and $\theta$ is
the polar angle with respect to the magnetic axis.

We take into account general relativistic photon transport for the
calculation of quantities that are measured by an observer at
infinity. We use the photon trajectories given by the Schwarzschild
metric, which is a good approximation for slowly rotating neutron
stars such as AXPs, and follow the method described by Pechenick,
Ftaclas, \& Cohen (1983) to compute light curves and spectra.

In our calculations, we explore the effects of varying the neutron
star properties as well as the orientation of the observer with
respect to the rotation axis of the star. The former parameter space
includes the surface temperature $T_{\rm eff}$, the magnetic field
strength $B$ on the magnetic pole, and the neutron star compactness,
described here by the relativity parameter $p
\equiv Rc^2/2GM$, where $R$ and $M$ are the radius and mass of the
neutron star, respectively. We consider effective temperatures and
magnetic field strengths in the ranges $0.3$~keV $\leq T_{\rm eff}
\leq 0.6$~keV and $10^{14}$~G $\leq B \leq 10^{15}$~G, respectively,
as suggested by the spectral and timing properties of AXPs. We also
consider a neutron-star compactness in a range such that $2 \le p \le
4$, which is allowed by current equations of state for a wide range of
neutron star masses and magnetic field strengths (see, e.g., Cardall,
Prakash, \& Lattimer 2001). In Figures~5-7, we assume a fiducial
neutron star radius of 10~km but discuss in \S4 the effects of
changing this value. In addition, the properties of the observed
radiation from the neutron star are affected by two angles, $\alpha$
and $\beta$, that specify, respectively, the position of the hot
region on the neutron star surface with respect to the rotation axis,
and of the observer with respect to the same axis.  We allow both of
these angles to vary between $0^\circ$ and $90^\circ$.

\subsection{Phase-Averaged Spectra}

\begin{figure}[t]
 \centerline{ \psfig{file=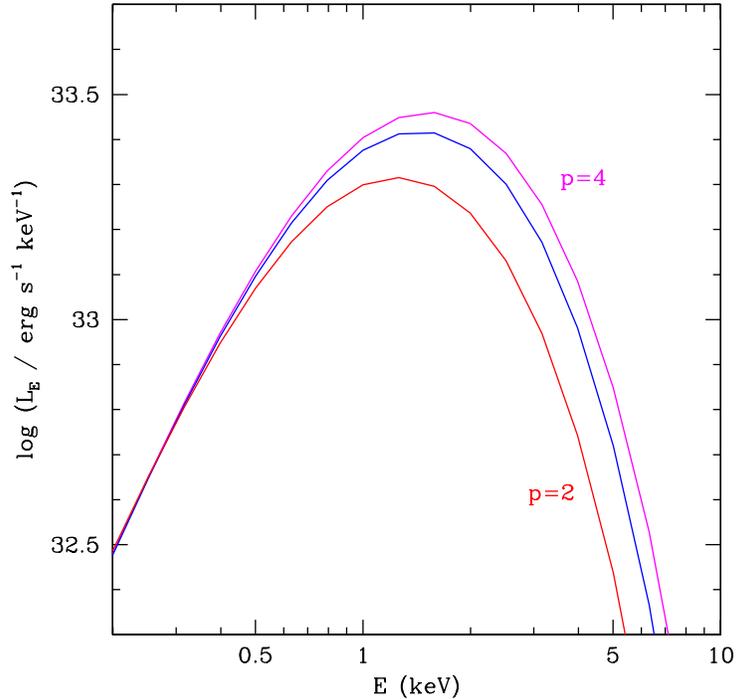,angle=0,width=10truecm} }
\figcaption[]{The phase-averaged spectrum, plotted as luminosity
measured by an observer at infinity at different photon energies, for
a neutron star with a magnetic field of $10^{14}$~G and two antipodal
hot emitting regions with effective temperature $T_{\rm eff}=0.5$~keV
and angular size $\rho=15^\circ$. Here, the two orientation angles
$\alpha$ and $\beta$ are taken to be $90^\circ$. \label{Fig:spec_av}}
\end{figure}

Figure~\ref{Fig:spec_av} shows the phase-averaged spectrum as measured
by an observer at infinity, for a neutron star of $B=10^{14}$~G with
two antipodal hot emitting regions of $T_{\rm eff}=0.5$~keV and
$\rho=15^\circ$, and different values of the relativity parameter
$p$. In this calculation, we assumed an orthogonal rotator (i.e.,
$\alpha=\beta=90^\circ$). However, when calculating the diagnostic
curves presented in \S4 and \S5, we take into account all possible
orientation angles. This is important because \"Ozel (2001b) shows
that the spectra depend sensitively on the choice of $\alpha$ and
$\beta$.

The spectra shown in Figure~\ref{Fig:spec_av} are broader than pure
blackbodies and show hard excesses. The color temperature
$T^\infty_{\rm c}$ measured at infinity of the best-fit blackbodies
are $T^\infty_{\rm c}\simeq 0.95\, T_{\rm eff}$ for $p=2$ and
$T^\infty_{\rm c}\simeq 1.15\,T_{\rm eff}$ for $p=4$, with $p=2$
yielding the spectrum with the highest gravitational redshift. In the
range of temperatures and magnetic field strengths considered here,
$T^\infty_{\rm c} \approx T_{\rm eff}$, as the significant color
correction arising from the atmosphere is partially offset by the
gravitational redshift. Therefore, the resulting color temperatures
are $T^\infty_{\rm c}\sim 0.4-0.6$~keV, comparable to those typical of
the blackbody components in the blackbody plus power-law decomposition
of the observed AXP spectra (see Table~\ref{Tab:BB}). We use this
entire range of values in the diagnostic plots discussed in \S4 (see
also \"Ozel 2001b for a detailed discussion and a table of color
temperatures including the models for $B=10^{15}$~G). Note that the
broad-band spectra presented here are consistent with those of Ho \&
Lai (2001) but differ from the results of Zane et al.\ (2001) who use
a local temperature-correction scheme that has poor convergence
properties and does not yield radiative-equilibrium solutions.

\subsection{Energy-Dependent Pulsed Fractions}

The pulse profiles and the corresponding pulsed fractions for a
thermally emitting neutron star measured by an observer at infinity
are determined by {\em (i)\/} the beaming of the radiation emerging
from the stellar surface, {\em (ii)\/} the temperature anisotropy on
the neutron star, and {\em (iii)\/} the general-relativistic photon
transport from the star to the observer (see, e.g., Page 1995).  For
the case of a thermally emitting magnetar, all three effects may
introduce a strong dependence of the pulsed fractions on photon
energy.

The beaming of radiation at the stellar surface of a strongly magnetic
neutron star has two local maxima: a narrow peak in the radial
direction, producing a pencil-like beam, and a broader one away from
the normal to the surface, producing a fan-like beam (see, e.g.,
M\'esz\'aros 1992; Pavlov et al 1994; \"Ozel 2001a). The relative
photon flux of the two peaks depends on both the photon energy and the
magnetic field strength, through the ratio $E/E_{\rm b}$, where $E$
and $E_{\rm b}$ are the photon and the electron cyclotron energies,
respectively. In particular, for strong magnetic fields ($B\gtrsim
10^{14}$~G), the fan beam dominates the emerging flux. In addition,
with increasing photon energy, this non-radial peak appears at larger
angles from the normal and becomes narrower. Such a radiation pattern
results in a pulsed fraction that typically increases strongly with photon
energy.

\begin{figure}[t]
 \centerline{ \psfig{file=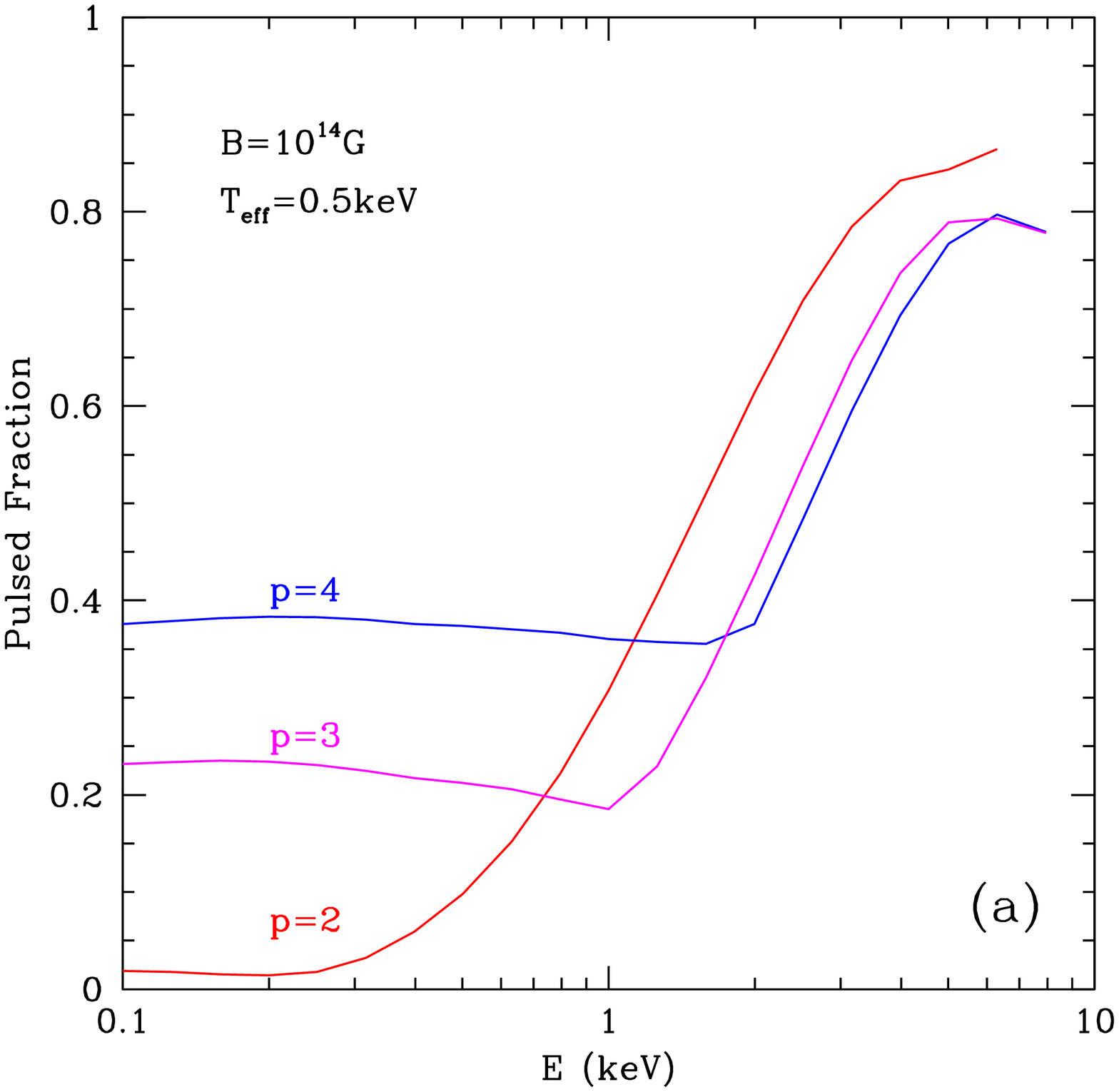,angle=0,width=8truecm} 
              \psfig{file=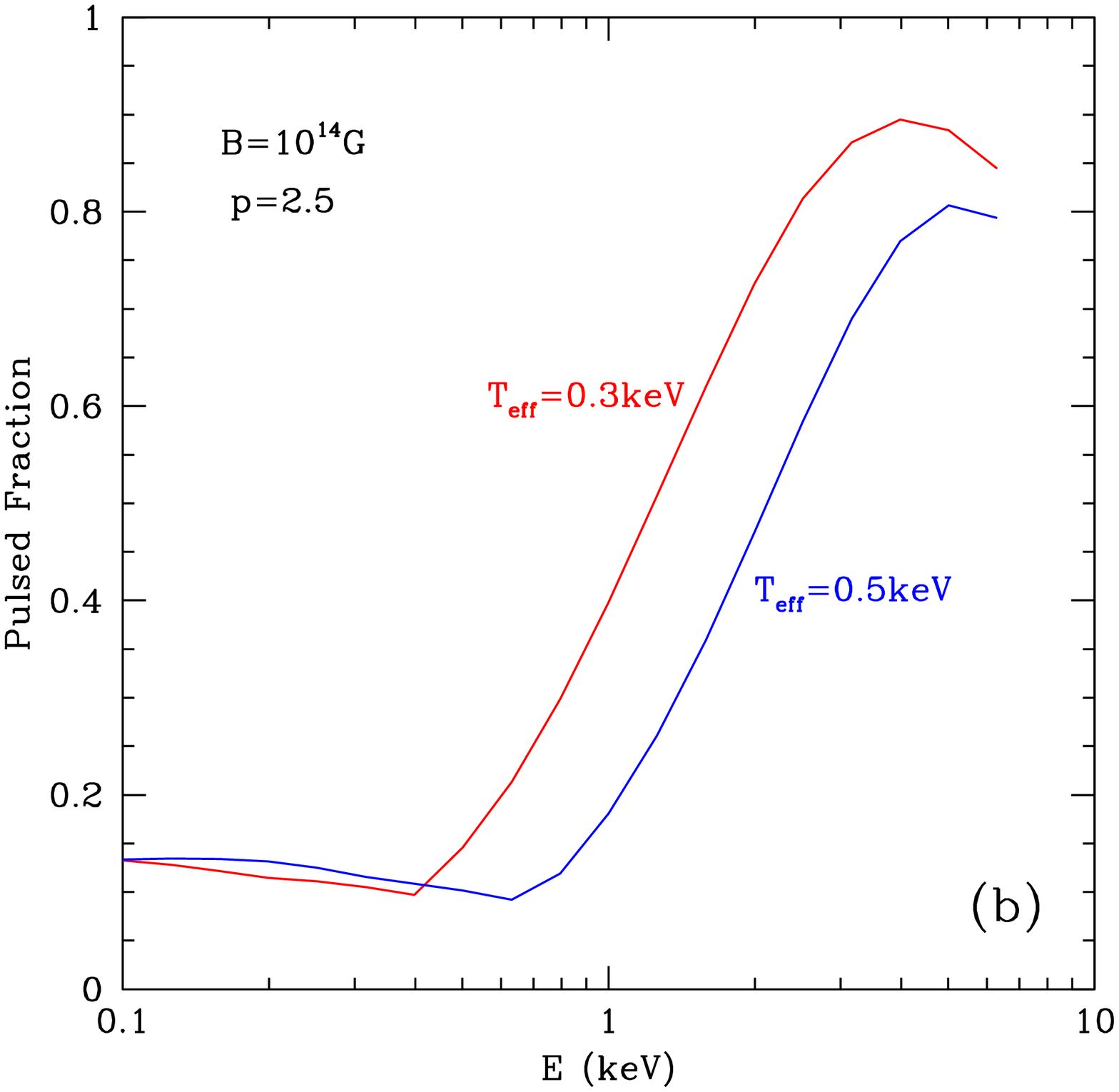,angle=0,width=8truecm} }
\figcaption[]{The predicted energy dependence of the pulsed fraction
for surface emission from a magnetar with two antipodal hot regions of
size $\rho=15^\circ$ and different values of the (a) neutron star
compactness and (b) effective temperature, for
$\alpha=\beta=90^\circ$. \label{Fig:pf_e}}
\end{figure}

Even in the absence of energy-dependent beaming, a temperature
anisotropy on the neutron star gives rise to an energy dependent
pulsed fraction (Page 1995); note that an anisotropy is always
required for the surface emission to be pulsed. This energy dependence
arises from an overall shift in the color temperature of the observed
spectrum during a pulse cycle, caused by the successive appearance of
hotter and colder regions along the line of sight. As a result, the
flux emitted in a narrow range of photon energies corresponding to the
peak of the phase-averaged spectrum shows little variation, while the
rest of the spectrum suffers large excursions. Therefore, the pulsed
fraction shows a minimum at a photon energy comparable to the peak of
the phase-averaged spectrum and increases strongly at higher energies
due to the exponential character of the spectrum.

Finally, the pulsed fraction measured by an observer at infinity
depends also on the curvature of the photon trajectories outside the
compact star. For radially peaked and energy independent beaming
patterns and for typical values of the neutron-star compactness,
general-relativistic effects spread the pulsar beam, suppressing the
observed variability (Pechenick et al.\ 1983). However, for the
fan-shaped beaming pattern relevant for strongly magnetized
atmospheres, general relativity may actually enhance the pulsed
fractions with increasing neutron-star compactness, by combining the
flux from two antipodal hot regions so that it peaks at a phase
$\pi/2$ away from the magnetic axis (\"Ozel 2001b). Moreover, even in
the case of radiation from a single emitting region, the two fan beams
at low photon energies may combine along the magnetic axis in a
similar way, depending on the orientation angles and the size of the
emitting region (also see below). Finally, depending on the
compactness of the neutron star, general relativistic redshifts alter
the peak of the phase-averaged spectrum and thus the pulsed fraction
at a given photon energy as measured by an observer at infinity.

Figure~\ref{Fig:pf_e} shows the energy dependence of the pulsed
fractions for strongly magnetic neutron stars with antipodal emission
geometry, for different values of the effective temperature and the
relativity parameter $p$. The results correspond to emission regions
with angular size $\rho=15^\circ$ and of uniform temperature $T_{\rm
eff}$. Note that considering temperature variations within the hot
caps increases the effects of temperature anisotropy. All curves have
the same characteristic shape. They are flat at low pulsed fractions
for photon energies below the peak of the phase-averaged spectrum,
show a sharp increase at higher energies, and flatten again at high
pulsed fractions. This prediction of a characteristic rise in the
pulsed fractions with photon energy is the first diagnostic that we
will compare with the observations discussed in \S2. The horizontal
displacement of the curves, along the energy axis, depends on the
effective temperature and the compactness of the neutron star.

\begin{figure}[h]
 \centerline{ \psfig{file=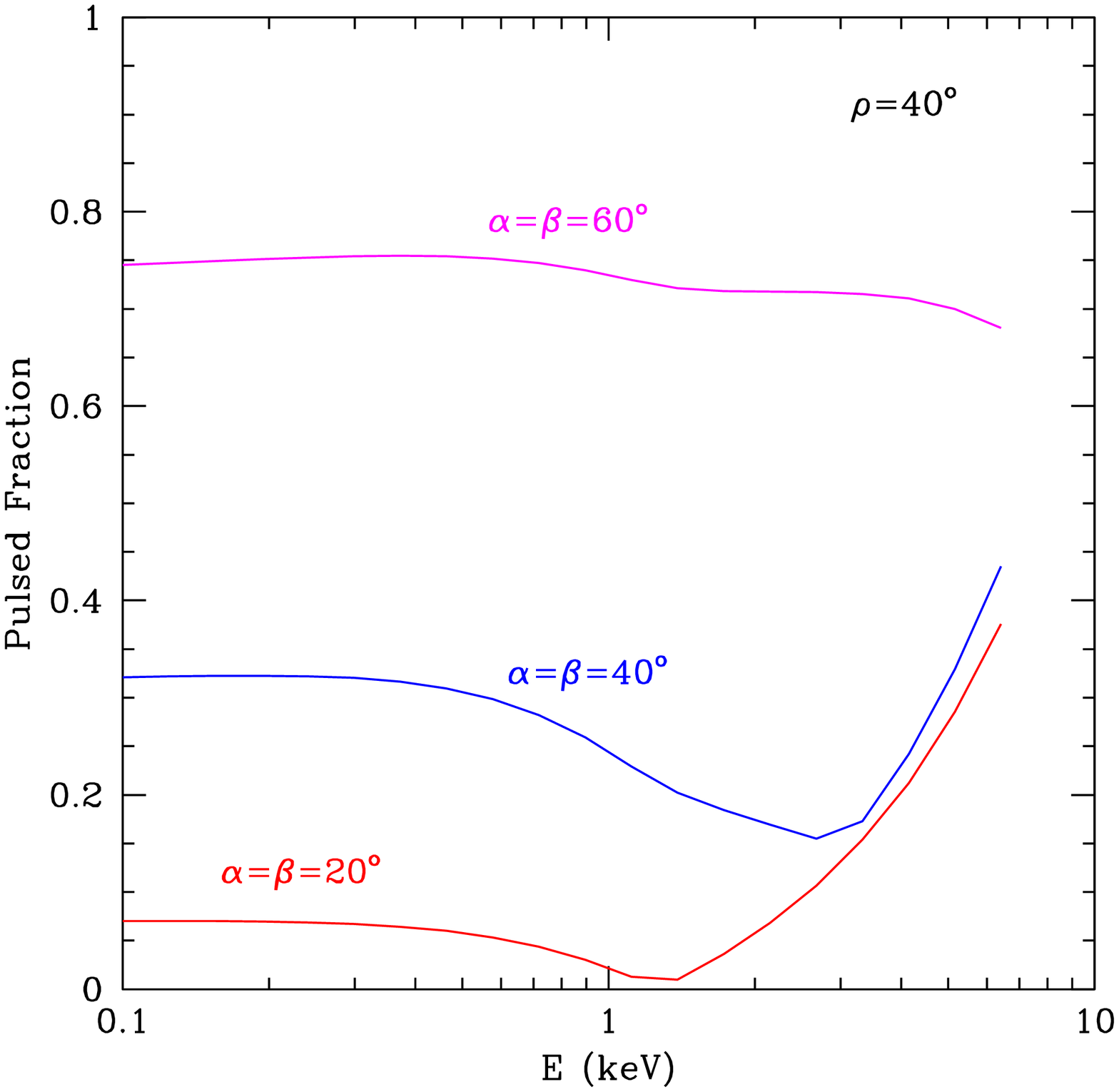,angle=0,width=8truecm}
              \psfig{file=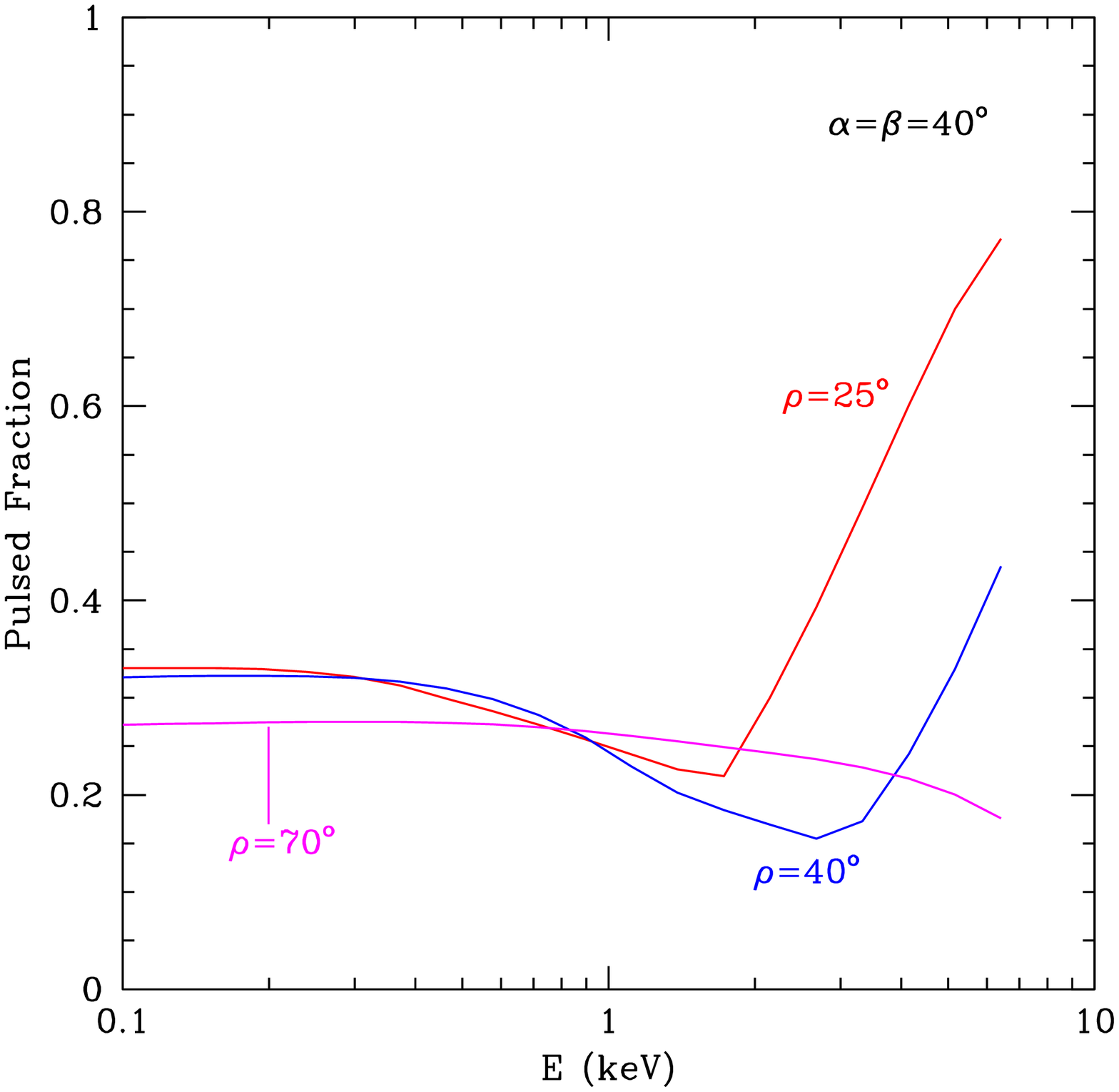,angle=0,width=8truecm} } 
\figcaption[]{Same as in Fig.~3 but for one hot emitting region and
a neutron star compactness corresponding to $p=2.5$. Curves in panel
(a) correspond to different values of the orientation angles for a
size of $\rho=40^\circ$. In panel (b) $\alpha=\beta=40^\circ$ and the
size $\rho$ of the hot region is varied. \label{Fig:pf_e1s}}
\end{figure}

Several other properties of the model affect the energy dependence of
the pulsed fractions. First, changing the magnetic field strength to
$B=10^{15}$~G leaves the pulsed fraction nearly unchanged at energies
$E \lesssim 5$~keV, but gives rise to a slightly larger value at
higher photon energies. This is because of the narrower fan beam of
the higher magnetic field case in this energy range. Second, changing
the orientation angles to values less than $\alpha=\beta=90^\circ$
suppresses the pulsed fraction at all photon energies but does not
affect this characteristic shape. Finally, the temperature
distribution given in equation~(\ref{eq:hh}) does not change the
overall characteristics of the energy dependence of the pulsed
fractions but leads to lower pulsed fractions overall because of the
larger effective emitting area of this model.

We also explore the possibility of a single hot emitting region on the
neutron-star surface, with the rest of the star assumed to produce no
radiation. This geometry is consistent with the presence of two peaks
in the pulse profiles of AXPs because of the non-radial beaming of the
surface emission from a magnetar (\"Ozel 2001b). The results are shown
in Figure~\ref{Fig:pf_e1s}.

While in general yielding a similar characteristically rising shape,
the pulsed fractions in this case show a much wider range of behavior
depending on the observer angle and the size of the emitting
region. In particular, the pulsed fraction can be flat or weakly
decreasing with photon energy, as shown in the two cases in
Figure~\ref{Fig:pf_e1s}. This is because for observer angles that
graze the emission cone, the significantly wider fan beams at low
photon energies may combine along the magnetic axis due to the general
relativistic effects and thus boost the flux observed at this pulse
phase. At high energies, however, the fan beams, which are narrower
and peak at larger angles away from the magnetic axis, remain distinct
and observable as two peaks in the pulse cycle (see also \"Ozel
2001b). The resulting pulse profiles can yield pulsed fractions that
are constant or weakly decreasing with photon energy
(Fig.~\ref{Fig:pf_e1s}a). Note that, when $\alpha\gtrsim 70^\circ$ and
$\beta \gtrsim 70^\circ$, the pulsed fractions at all photon energies
are close to unity.

A similar effect can be obtained by increasing the size of the
emitting region (Fig.~\ref{Fig:pf_e1s}b). While decreasing the overall
pulsed fraction, a large enough size of the emitting region also gives
rise to weaker energy dependences of the pulsed fractions for the
reasons discussed above. Note that this very weak energy dependence of
the pulsed fraction shown in Figure~\ref{Fig:pf_e1s} can be obtained
for a significant region of the parameter space.

\subsection{Broad-Band Variability}

The pulsed fractions of thermally emitting magnetars typically
increase with photon energy (\S3.3). Therefore, the broad-band
variability properties, which are relevant for comparison of
theoretical models to data, depend on the chosen range of photon
energies as well as on all the effects discussed in \S3.3. In
addition, interstellar extinction preferentially reduces the flux of
low-energy photons that reach the observer. Thus, for the pulsed
fraction that increases with photon energy, this preferential
attenuation may increase the pulsed fraction measured over a wide
energy range (Page 1995).
 
\begin{figure}[t]
 \centerline{ \psfig{file=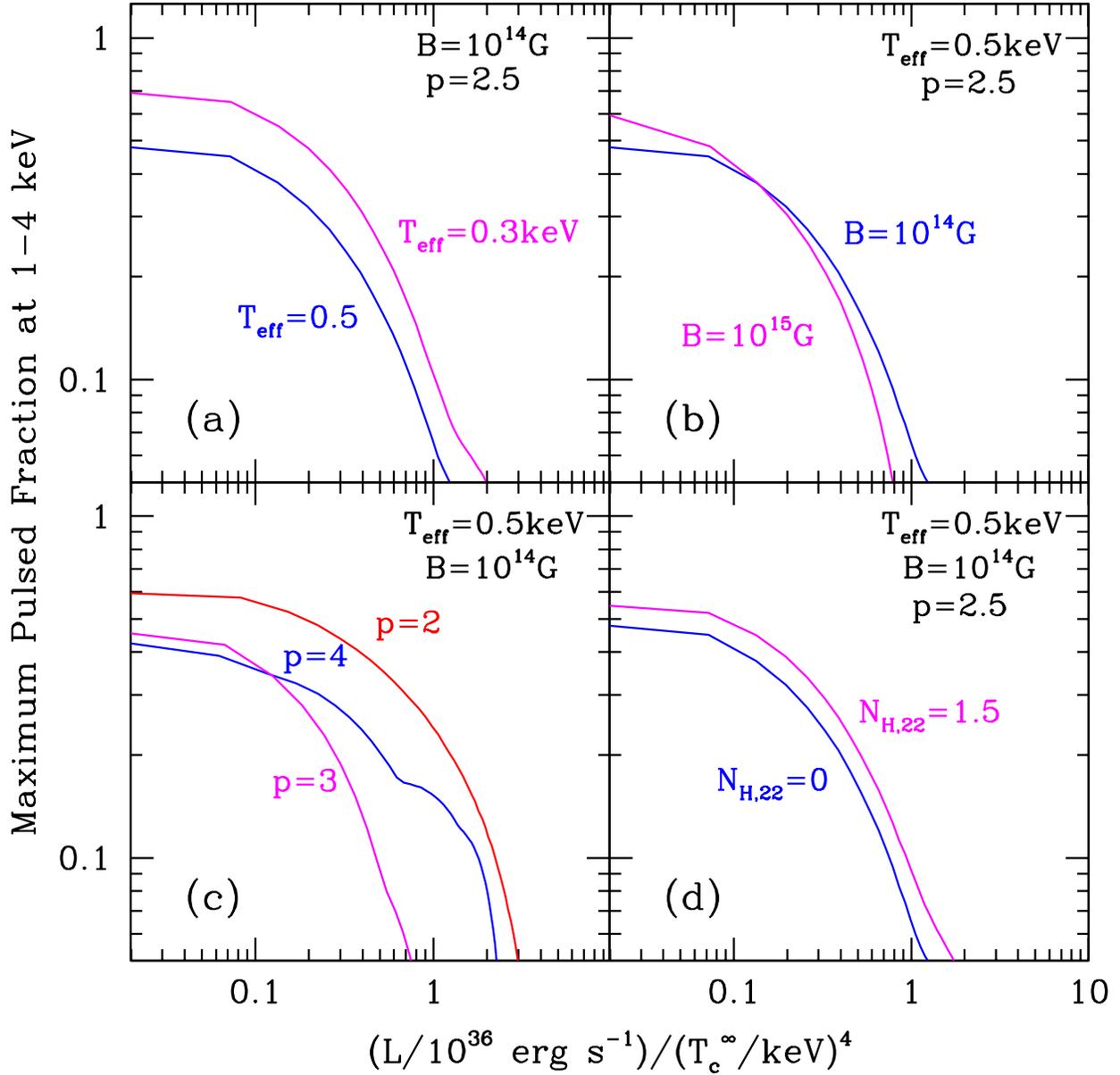,angle=0,width=17truecm}}
\figcaption[]{The maximum pulsed fractions in the $1-4$~keV range plotted 
against the ratio of the luminosity to the fourth power of the color
temperature, as inferred by an observer at infinity, for different
values of the (a) effective temperature, (b) magnetic field strength,
(c) compactness of the neutron star, and (d) hydrogen column to the
source in units of $10^{22}~{\rm cm}^{-2}$.  For each curve, regions
below and to the left of the curve are allowed, whereas the regions
above and to the right are excluded for any choice of $\alpha$ and
$\beta$. \label{Fig:pl_the}}
\end{figure}

Most importantly, the pulsed fraction depends on the number and size
of the hot emitting regions on the neutron star, which at the same
time determine the total luminosity. For a given effective
temperature, the luminosity decreases, while the pulsed fraction
generally increases with decreasing size of the emitting
region. Therefore, for any energy range, there exists a maximum pulsed
fraction that can be attained for a given luminosity and for any
combination of the orientation angles $\alpha$ and $\beta$.
Figure~\ref{Fig:pl_the} shows the maximal curves for different values
of the model parameters. Note that in all cases, the orthogonal
rotator geometry ($\alpha=\beta=90^\circ$) gives rise to the maximal
curve. No thermally emitting neutron star is allowed to lie above and
to the right of these limiting curves, while the region below and to
the left is permitted, since these smaller values of luminosity and
pulsed fraction can be obtained for smaller values of the orientation
angles without changing the other model parameters. The comparison of
the allowed regions for the thermally emitting magnetar model to the
observations of AXPs is the second diagnostic test that we perform in
\S4.

All quantities plotted in Figure~\ref{Fig:pl_the} can be directly
determined by observations. Specifically, the abscissa is the ratio of
the energy-integrated luminosity $L$ of the thermal emission to the
fourth power of the blackbody temperature $T^\infty_{\rm c}$ that best
reproduces its spectrum. The ordinate is the pulsed fraction of the
photon flux measured in the $1-4$~keV energy range. This is the energy
range in which the blackbody contribution to the photon flux is
maximum and for which good determinations of pulsed fractions exist
for all five AXPs considered here (see Fig.~\ref{Fig:bbody}).

We assume a fiducial neutron star radius of 10~km in calculating the
quantity plotted in the abscissa of Figures~5-7. Considering, e.g., a
15~km radius would result in a $(R/10~{\rm km})^2=(1.5)^2$ increase in
the total luminosity and thus would shift the curves to the right by
this factor. The dependence of the curves on $T_{\rm eff}$, $B$, and
$p$, shown in the first three panels, follows directly from the
discussion of the energy-dependent pulsed fraction presented in \S
3.3. The last panel shows the dependence of the maximum pulsed
fraction on the hydrogen column $N_{\rm H}$ that describes the effect
of interstellar extinction for the range of values measured from
fitting the X-ray spectra of AXPs (see Table~\ref{Tab:BB}). The
increase of the $1-4$~keV pulsed fraction with $N_{\rm H}$ is marginal
because of the narrow photon-energy range used as well as the
relatively large values of $T^\infty_{\rm c}$ and small values of
$N_{\rm H}$ indicated by observations (see Table~3). The large effect
of interstellar extinction on pulsed fractions calculated by Perna et
al.\ (2000) can only be obtained for significantly lower values of the
color temperature ($\sim 0.1$~keV) and higher values of the hydrogen
column density ($\sim 10^{23}$~cm$^{-2}$), which are inconsistent with
observations (\S2).

\section{CONSTRAINTS ON THERMAL EMISSION MODELS FROM DATA}

In this section we compare the predictions of the different thermally
emitting magnetar models to the observations of AXPs in order to
determine those properties that can be accounted for in the models.
In particular, we discuss {\em (i)\/} the energy dependence of the
pulsed fractions and {\em (ii)\/} the maximum pulsed fractions and
thermal luminosities that can be simultaneously produced by a
magnetar.

The weak energy dependence of the observed AXP pulsed fractions in the
$1-10$~keV range is at odds with the characteristic steeply rising
pulsed fractions predicted by the models with two antipodal emitting
regions. As Figure~\ref{Fig:pf_e} shows, this steep increase of the
pulsed fraction typically occurs in the $1-5$~keV energy band for the
wide range of model parameters considered here, and does not depend on
the effects of interstellar extinction. However, the detector response
may play a significant role in giving rise to the observed weak
dependence (Page 1995). This is primarily caused by the fact that the
detector response is non-local in photon energy and a fraction of the
counts that are detected at low-energy channels correspond to high
energy photons. This may increase the pulsed fraction measured at low
energies. 

For the case of the single emitting region, a range of orientation
angles combined with large emitting areas gives rise to pulsed
fractions that depend very weakly on photon energy (\S3.3). Therefore,
for this range of angles and hot region sizes, the model with a single
emission region can reproduce the weak energy dependence observed in
the pulsed fractions of AXPs. For other angles, the steeply rising
behavior is present for photon energies $E \gtrsim 1$~keV (see
Fig.~4).  The range of parameters that yields the observed weak
dependence may be affected by the detector response and needs to be
explored in more detail.  X-ray observations with high spatial and
energy resolutions and good coverage of the $\sim 0.1-5$~keV energy
band made with {\em Chandra} and {\em XMM-Newton\/} can provide the
conclusive test of this prediction.

\begin{figure}[t]
 \centerline{ \psfig{file=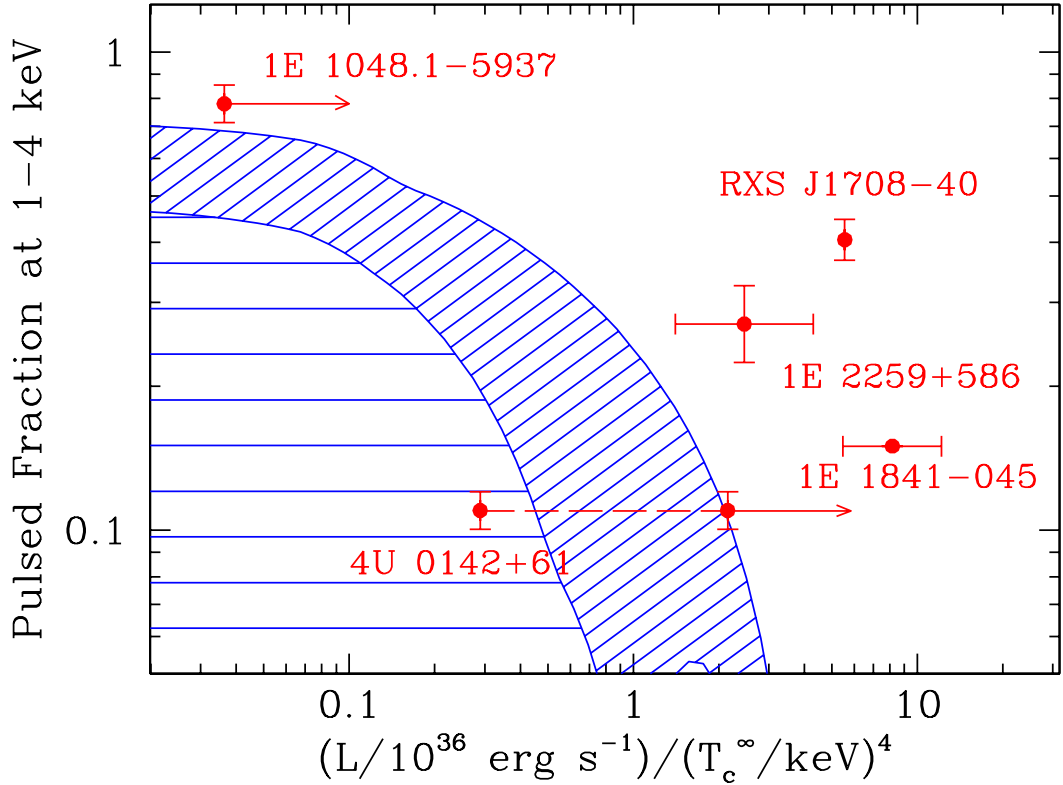,angle=-90,width=15truecm}}
\figcaption[]{Comparison of the AXP data to the maximal predictions
of the thermally emitting magnetar model with two antipodal emitting
regions. The diagonally shaded band shows the maximum pulsed fraction
as a function of the ratio of luminosity to the fourth power of the
color temperature, and spans the entire range of curves shown in
Figure~5 for $\alpha=\beta=90^\circ$. The horizontally shaded region
is also allowed for smaller orientation angles. The right arrows on
the data points indicate conservative lower limits, while the error
bars represent an allowed range of values. Most of the sources lie
outside the allowed region of the diagram. \label{Fig:2spot}}
\end{figure}

In Figure~\ref{Fig:2spot} we compare to observations the maximum
pulsed fractions allowed by magnetar models, for neutron stars of
different luminosities and color temperatures and with two antipodal
emitting regions. The diagonally shaded band spans the range of all
the maximal curves plotted in Figure~\ref{Fig:pl_the} and corresponds
to the entire range of neutron-star parameters that are consistent
with the spectral data. Note that this band is obtained for
$\alpha=\beta=90^\circ$ which yields the maximum pulsed fraction.
Varying the orientation angles results in bands that lie below the
maximal one and cover the whole horizontally shaded region. Therefore,
to be consistent with a thermally emitting magnetar model, all sources
must lie within the shaded regions.

The data points are a synthesis of the various observed quantities
summarized in Tables~$2-4$. In the case of 1E~2259$+$586 and
1E~1841$-$045, the horizontal error bars correspond to the
uncertainties in the parameters of the spectral fits and the most
conservative range of distance estimates to the sources. For
1E~1048.0$-$5937 and 4U~0142$+$61, the data points represent the
conservative lower limits on the distance estimates discussed in \S
2.3. (Note that the two such limits for the latter source arise from
the uncertain effect on the distance estimate of a molecular cloud
along the line of sight). Finally, the luminosity of RXS~J1708$-$40
has been inferred for the adopted fiducial distance.

For the antipodal emission geometry considered in
Figure~\ref{Fig:2spot}, all data points but one lie very far outside
the region allowed for thermal emission from a magnetar. In addition,
for orientation angles different from those of an orthogonal rotator,
the discrepancy becomes rapidly larger, owing to the quick drop in the
pulsed fraction with decreasing orientation angles in this antipodal
configuration. We also emphasize that the regions outside of the hot
caps are assumed not to contribute at all to the emission from the
neutron star, and a more realistic temperature distribution would
produce even lower pulsed fractions.  Finally, for the temperature
profile of a cooling magnetar given by Equation~\ref{eq:hh}, the
maximum pulsed fraction attained for a given luminosity is
significantly below the diagonally shaded region. We therefore
conclude that the properties of AXPs cannot be accounted for by a
thermally emitting magnetar model with two identical antipodal
emitting regions or one with the temperature profile of
Equation~\ref{eq:hh}. Note that considering a larger neutron star
radius (see \S3.4 ) would still not be enough to account for the data
points and does not alter the results.

\begin{figure}[t]
 \centerline{ \psfig{file=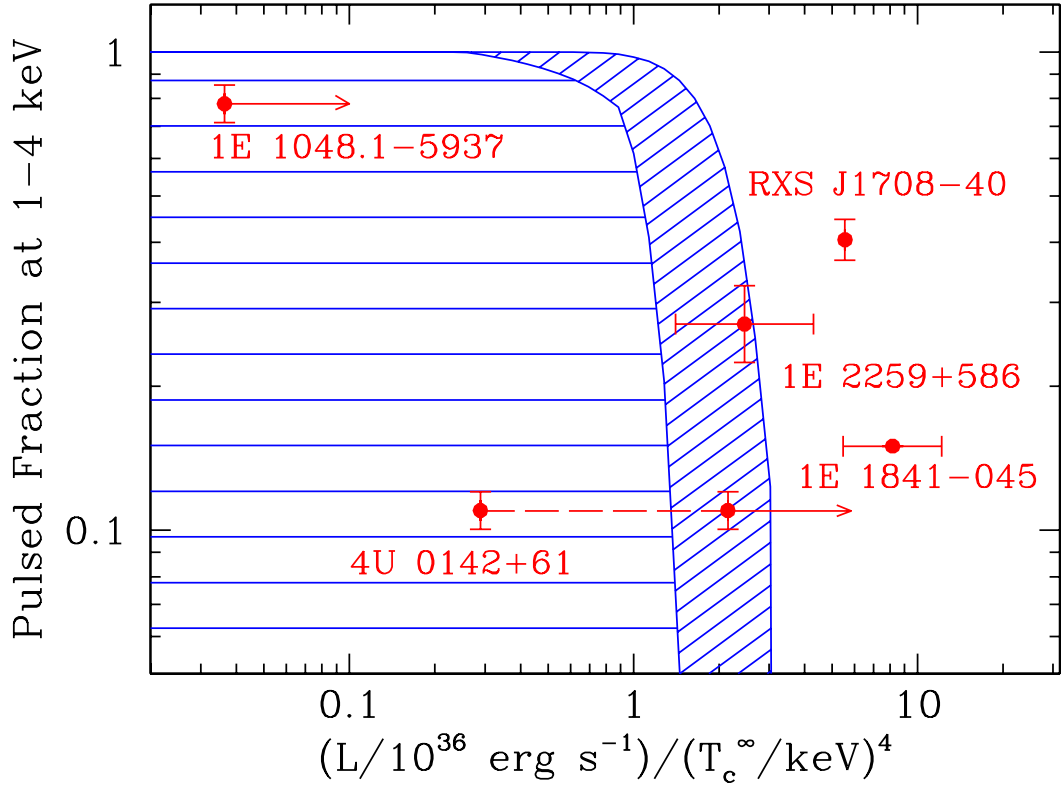,angle=-90,width=15truecm} }
\figcaption[]{Same as in Fig.~6, but for a single emitting region on the 
 neutron star. Most data points lie within the allowed
 region. \label{Fig:1spot}}
\end{figure}

We then consider a single emitting region on the neutron star surface.
The result is shown in Figure~\ref{Fig:1spot}. The band of maximal
curves is consistent with all but two data points.  For these two
sources to be reconciled with the theoretical models, the distance to
RXS~J1708$-$40 needs to be revised downwards by $\sim 30\%$ of the
current best estimate, and the color temperature and blackbody
luminosity measured for 1E~1841$-$045 needs to be corrected for the
effects of the hard spectral tail (see \S2). We emphasize that in most
cases, even though the maximal band is obtained for an orthogonal
rotator, the consistency of the data with the models is not limited to
a narrow range of orientations close to $\alpha=\beta=90^\circ$ but is
possible for a reasonable fraction of the parameter space. Finally, we
note that a single emitting region configuration may be realized in
the case of, e.g., largely asymmetric magnetic activity between the
magnetic poles, an offset dipole magnetic field, or even for a
non-dipolar geometry. However, the detailed properties of a physical
mechanism that may lead to such an anisotropy have not yet been
explored.

\section{CONCLUSIONS}

We have compared the detailed models of thermal emission from the
surface of a magnetar (\"Ozel 2001a, b) to the observed properties of
AXPs. This comparison is relevant if the inferred blackbody component
in the X-ray spectra of AXPs arises from surface emission. We have
included in our model calculations the effects of a strongly
magnetized atmosphere on the angle- and energy-dependence of the
surface emission and the general relativistic photon transport to an
observer at infinity. We have focused on the combined spectral and
variability properties of the sources and specifically on their
luminosities and pulsed fractions.

We have found that the large pulsed fractions combined with the high
inferred luminosities of AXPs cannot be accounted for by surface
emission from a magnetar with two antipodal hot regions or a
temperature distribution characteristic of a magnetic dipole. Models
with a single hot emitting region can reproduce the observations, if
we allow for systematic uncertainties in the luminosity estimates of
two of the sources, RXS~J1708$-$40 and 1E~1841$-$045. Conversely, the
maximum allowed distances to all the sources are constrained within
this model and the diagnostic can be used as an independent distance
measurement if further study shows this model to be relevant for AXPs.

An even more detailed test of the models can be performed by comparing
the predicted energy dependence of the pulsed fractions to
observations. Current data appear to be inconsistent with the
predictions of the model with antipodal emission geometry, but can be
reproduced for a range of orientation angles and sizes of the emitting
areas if the emission is localized to a single region on the
surface. Upcoming observations of these sources with the {\em Chandra}
and {\em XMM-Newton} observatories will yield the high spatial- and
energy-resolution data that are required for this diagnostic test.

\acknowledgements

We thank Ramesh Narayan for many useful discussions.  F.\,\"O.\ and
D.\,P.\ thank the Physics Department at McGill University for their
hospitality and Fotis Gavriil for his help with XSPEC. This work was
supported in part by NSF Grant AST-9820686 (F.\,\"O.), by NASA LTSA
grants NAG 5-9184 (D.\,P.) and NAG5-8063 (V.\,M.\,K.), by an Alfred
P.\,Sloan fellowship (V.\,M.\,K.), and by NSERC Research Grant
RGPIN228738-00 (V.\,M.\,K.).

\newpage

\begin{deluxetable}{lcccc}
\tablewidth{450pt} 
\tablecaption{AXP TIMING PROPERTIES}
\tablehead{Source Name & $P$~(s) & $\dot{P}$~($10^{-12}$~s$\,$s$^{-1}$) &
   $B_{\rm p}$ ($10^{14}$~G)\tablenotemark{a} &
References\tablenotemark{b}} 
\startdata
1E~1048.1--5937 & 6.45 & 8.5--38.1 & 2.4--5& 1\\ 
1E~2259$+$586   & 6.98 & 0.49 & 0.6 & 2\\
4U~0142$+$61    & 8.69 & $\sim 2$ & 1.3 & 3\\
RXS~J1708$-$40 & 11.0 & $\sim 19.0$ & 4.6 & 2\\
1E~1841$-$045   & 11.8 & $\sim 47.3$ & 7.5 & 4
\enddata

\tablenotetext{a}{Inferred dipole magnetic field at the stellar pole
for an orthogonal rotator.}

\tablenotetext{b}{References: 1.\ Kaspi et al.\ 2001; 2.\ Kaspi, 
Chakrabarty, \& Steinberger 1999; 3.\ Israel et al.\ 1999; 4.\
Vasisht \& Gotthelf 1997}

\label{Tab:B} 
\end{deluxetable}

\begin{deluxetable}{lcc}
\tablewidth{450pt} 
\tablecaption{AXP PULSED FRACTIONS\tablenotemark{a}}
\tablehead{Source Name & Energy Range (keV) & Pulsed Fraction}
\startdata
1E~1048.1--5937 & 1.0--1.5 & 0.76 $\pm$ 0.08 \\
                & 1.5--2.0 & 0.74 $\pm$ 0.05 \\
       		& 2.0--2.5 & 0.77 $\pm$ 0.04 \\
 		& 2.5--4.0 & 0.84 $\pm$ 0.03 \\
 		& 4.0--6.0 & 0.79 $\pm$ 0.05 \\
	 	& 6.0--10.0 & 0.54 $\pm$ 0.12\\
\hline
1E~2259$+$586   & 0.2--2.2(?) & 0.27 $\pm$ 0.02\\
		& 0.5--2.0 & 0.35 $\pm$ 0.14\\
 		& 2.5--3.5 & 0.27 $\pm$ 0.05\\ 
\hline
4U~0142$+$61    & 0.5--1.5 & 0.12 $\pm$ 0.01\\
		& 1.5--4.0 & 0.11 $\pm$ 0.01\\
		& 4.0--10.0 & 0.17 $\pm$ 0.02\\
\hline 
RXS~J1708$-$40 & 0.8--2.0 & 0.39 $\pm$ 0.04\\
			& 2.0--4.0 & 0.42 $\pm$ 0.04\\
			& 4.0--10.0 & 0.45 $\pm$ 0.15\\ 
\hline
1E~1841$-$045 & 1.0--10.0 (?) & $\sim 0.15$\\ 
\enddata

\tablenotetext{a}{See text for references}

\label{Tab:PF} 
\end{deluxetable}

\begin{deluxetable}{lcccl}
\tablewidth{500pt} 
\tablecaption{AXP BLACKBODY SPECTRA}
\tablehead{Source Name & $T_{\rm BB}$ (keV)\tablenotemark{a} & 
   $R_{\rm BB}$ (km)\tablenotemark{b} & $N_{\rm H} (10^{22}$~cm$^{-2}$) 
   & Notes\tablenotemark{c}}
\startdata
1E~1048.1--5937 & 0.64 $\pm$ 0.01& (0.59 $\pm$ 0.02)$d_{\rm 3kpc}$ & 
   0.45 $\pm$ 0.10 & 1 (solid)\\
		& 0.52 $\pm$ 0.02& (0.81 $\pm$ 0.07)$d_{\rm 3kpc}$ & 
   0.55 $\pm$ 0.17 & 2 (short-dashed)\\ 
		& 0.56 $\pm$ 0.06& (0.65 $\pm$ 0.16)$d_{\rm 3kpc}$ & 
   1.21 $\pm$ 0.24 & 3 (dotted)\\ 
\hline
1E~2259$+$586   & 0.44 $\pm$ 0.01& (3.3 $\pm$ 0.3)$d_{\rm 4kpc}$ & 
   0.87 $\pm$ 0.05 & 4\\
\hline
4U~0142$+$61    & 0.39 $\pm$ 0.01 & $\simeq$2.4$d_{\rm 1kpc}$ 
  & 0.95 $\pm$ 0.04 & 5 (solid)\\
		& 0.38 $\pm$ 0.01 & (2.2 $\pm$ 0.2)$d_{\rm 1kpc}$ 
  & 1.17 $\pm$ 0.04 & 3 (Aug 98--dotted)\\
		& 0.42 $\pm$ 0.02 & (1.5 $\pm$ 0.2)$d_{\rm 1kpc}$ 
  & 1.11 $\pm$ 0.07 & 6 (obs A--short-dashed)\\
		& 0.36 $\pm$ 0.01 & (1.5 $\pm$ 0.5)$d_{\rm 1kpc}$ 
  & 0.98 $\pm$ 0.06 & 6 (obs D--long-dashed)\\
\hline 
RXS~J1708$-$40 & 0.41 $\pm$ 0.03 & (8.2 $\pm$ 1.6)$d_{\rm 10kpc}$ 
   & 1.42 $\pm$ 0.02 & 7\\
\hline
1E~1841$-$045 & 0.65 $\pm$ 0.05 & $\simeq 8$$d_{\rm 7kpc}$ &
1.2 $\pm$ 0.2 & 8 (only blackbody fit)
\enddata

\tablenotetext{a}{Fitted blackbody temperature}
\tablenotetext{b}{Apparent blackbody radius assuming isotropic
emission from a sphere and a specific distance.}
\tablenotetext{c}{Line-style for Figure~1 and references: (1)
Oosterbroek et al.\ 1998; (2) Corbet \& Mihara 1997; (3) Paul et al.\ 2000;
(4) Parmar et al.\ 1998; (5) White et al.\ 1996; (6) Israel et al.\ 1999;
(7) Sugizaki et al.\ 1997; (8) Gotthelf \& Vasisht 1997; Gotthelf,
Vasisht, \& Dotani 1999.}

\label{Tab:BB} 
\end{deluxetable}

\newpage

\begin{deluxetable}{lcl}
\tablewidth{450pt} 
\tablecaption{AXP DISTANCES AND LUMINOSITIES}
\tablehead{Source Name & Distance (kpc) & $L_{\rm BB}$ 
(erg~s$^{-1}$)\tablenotemark{a}}
\startdata
1E~1048.1--5937 & $\gtrsim 2.7$ & $\gtrsim 6\times 10^{33}$\\
1E~2259$+$586   & $4-7$ & $5\times 10^{34}-2\times 10^{35}$\\
4U~0142$+$61    & $\gtrsim 1.0$ or $\gtrsim 2.7$ & 
        $\gtrsim 7\times 10^{33}$ or $\gtrsim 5\times 10^{34}$\\
RXS~J1708$-$40 & $\sim 8$ & $\sim 2\times 10^{35}$\\
1E~1841$-$045  & $5.7-8.5$ & $10^{36}-2\times 10^{36}$
\enddata

\tablenotetext{a}{Bolometric blackbody luminosity inferred from
phase-averaged spectra (see Table~3).}

\label{Tab:L} 
\end{deluxetable}

\end{document}